\documentclass[12pt,epsfig]{article}
\usepackage{epsfig}
\begin{document}
\baselineskip=15pt
%
\newcommand{\bc}{\begin{center}}
\newcommand{\ec}{\end{center}}
\newcommand{\be}{\begin{equation}}
\newcommand{\ee}{\end{equation}}
\newcommand{\PP}{\mbox{$\Psi$}}
\newcommand{\wdg}{\mbox{$\wedge$}}
\newcommand{\bq}{\begin{eqnarray}}
\newcommand{\eq}{\end{eqnarray}}
\newcommand{\del}{\mbox{$\delta$}}
\newcommand{\AC}{\mbox{${\bf {\cal A}}$}}
\newcommand{\BA}{\mbox{${\bf A}$}}
\newcommand{\BB}{\mbox{${\bf B}$}}
\newcommand{\DC}{\mbox{${\bf {\cal D}}$}}
\newcommand{\xx}{\mbox{${\bf x}$}}
\newcommand{\yy}{\mbox{${\bf y}$}}
\newcommand{\Lam}{\mbox{${\Lambda}$}}
\newcommand{\lam}{\mbox{${\lambda}$}}
\newcommand{\eps}{\mbox{$\epsilon$}}
\newcommand{\vol}{\mbox{$d^{3}{\bf x}$}}
\newcommand{\Sc}{Schr\"odinger\ }
\newcommand{\p}{\mbox{$\varphi$}}
\newcommand{\der}{\partial}
\begin{titlepage}
\vskip1in 
\begin{center}
{\Large{\bf Pseudoscalar mesons in asymmetric matter}}
\end{center}
\vskip1in
\begin{center}
{\large {\bf Maria C. Ruivo}$^\dagger$,
{\bf Pedro Costa}$^\dagger$ and {\bf C\'{e}lia A. de Sousa}$^\dagger$}\\
\vskip0.5cm
$^\dagger$ Centro de F\'{\i}sica Te\' orica, Departamento de F\'{\i}sica\\
Universidade de Coimbra \\P-3004-516 Coimbra, Portugal\\ 
\end{center}

\vskip0.5cm
\begin{center}
{\it maria@teor.fis.uc.pt, pcosta@fteor5.fis.uc.pt, celia@teor.fis.uc.pt}
\end{center}
\vskip1in
\begin{abstract}
\noindent
The behavior of kaons and pions in hot non strange quark matter, simulating
neutron matter, is investigated within the SU(3) Nambu-Jona-Lasinio [NJL]
and in the Enlarged Nambu-Jona-Lasinio [ENJL] (including vector
pseudo-vector interaction) models. At zero temperature, it is found that in
the NJL model, where the phase transition is first order, low energy modes
with $K^-, \pi^+$ quantum numbers, which are particle-hole excitations of
the Fermi sea, appear. Such modes are not found in the ENJL model and in NJL
at finite temperatures. The increasing temperature has also the effect of
reducing the splitting between the charge multiplets.
\end{abstract}

\end{titlepage}

\section{\bf Introduction}

During the last few years, major experimental and theoretical efforts have
been dedicated to heavy-ion collisions aiming at understanding the
properties of hot and dense matter and looking for signatures of phase
transitions to the quark-gluon plasma. As a matter of fact, it is believed
that, at critical values of the density, $\rho_c$, and/or temperature, $T_c$%
, the system undergoes a phase transition, the QCD vacuum being then
des\-cri\-bed by a weakly interacting gas of quarks and gluons, with
restored chiral symmetry.

The nature of the phase transition is an important issue nowadays. Lattice
simulations \cite{Kanaya} provide information at zero density and finite
temperature but for finite densities no firm lattice results are available
and most of our knowledge comes from model calculations. The
Nambu-Jona-Lasinio\cite{NJL} [NJL] type models have been extensively used
over the past years to describe low energy features of hadrons and also to
investigate restoration of chiral symmetry with temperature or density \cite
{Hatsuda94,Ruivo,SousaRuivo,RSP,Hiller}.

Recently, it was shown by Buballa\cite{buballa} that, with a convenient
parameterization, the SU(2) and SU(3) NJL models exhibit a first order phase
transition, the system being in a mixed phase between $\rho\,=\,0$ and $%
\rho\,=\,\rho_c$, the energy per particle having an absolute minimum at $%
\rho\,=\,\rho_c$. This suggests an interpretation of the model within the
philosophy of the MIT bag model. The system has two phases, one consisting
of droplets of quarks of high density and low mass surrounded by a non
trivial vacuum and the other one consisting of a quark phase of restored
chiral symmetry. Similar concepts appear in NJL inspired models including
form factors \cite{raja,david}. The physical meaning of the mesonic
excitations in the medium within this interpretation of the model is an
interesting subject that will be analyzed in this paper.

The possible modifications of meson properties in the medium is an important
issue nowadays.  The study of pseudoscalar mesons, such as kaons and pions,
is particularly interesting, since, due to their Goldstone boson nature,
they are intimately associated with the breaking of chiral symmetry. Since
the work of Kaplan and Nelson \cite{Kaplan}, the study of medium effects on
these mesons in flavor asymmetric media attracted a lot of attention.
Indeed, the charge multiplets of those mesons, that are degenerated in
vacuum or in symmetric matter, were predicted to have a splitting in flavor
asymmetric matter. In particular, as the density increases, there would be
an increase of the mass of $K^+$ and a decrease of the mass of $K^-$; a
similar effect would occur for $\pi^-$ and $\pi^+$ in neutron matter. The
mass decrease of one of the multiplets raises, naturally, the issue of meson
condensation, a topic specially relevant to Astrophysics.

Most theoretical approaches dealing with  kaons in flavor asymmetric media,
predict a slight raising of the $K^+$ mass and a pronounced lowering of the $%
K^-$ mass \cite{Lutz94,RSP,Cassing,Schaffner}, a conclusion which is
supported by the analysis of data on kaonic atoms \cite{Friedmann94}.
Experimental results at GSI seem to be compatible with  this scenario \cite
{Schr94,Herrman,Barth}.

Studies on pions in asymmetric medium are mainly related with the problem of
the $u-d$ asymmetry in a nucleon sea rich in neutrons. Such flavor asymmetry
has been established in SIS and DY experiments and theoretical studies show
that there is a significant difference in $\pi^+\,\,,\pi^-$ distribution
functions in neutron rich matter \cite{peng}.

From the theoretical point of view, the driving mechanism for the mass
splitting is attributed mainly to the selective effects of the Pauli
principle, although, in the case of $K^-$, the interaction with the $\Lambda
(1405)$ resonance plays an important role as well. In the study of the
effects of the medium on hadronic behavior, one should have in mind that the
medium is a complex system, where a great variety of medium particle-hole
excitations occur, some of them with the same quantum numbers of the hadrons
under study; the interplay of all these excitations might play a significant
role in the modifications of hadron properties. In previous works we have
established, within the framework of NJL models, the presence, in flavor
asymmetric media, of low energy pseudoscalar modes, which are excitation of
the Fermi sea \cite{RuivoSousa,SousaRuivo}. The combined effect of density
and temperature, as well as the effect of vector interaction, was discussed
for the case of kaons in symmetric nuclear matter without strange quarks 
\cite{RSP,Ruivo}.

This paper addresses the following points: a) analyzes of the phase
transition with density and temperature in neutron matter in the $SU(3)$ NJL
model with two different parameterizations and within the ENJL model ; b)
behavior of kaonic and pionic excitations in these models and discussion of
the meaning of the Fermi sea excitations, in connection with the nature of
the phase transition; c) combined effect of density and temperature.

\section*{Formalism}

We work in a flavor $SU(3)$ NJL type model with scalar-pseudoscalar and
vector-pseudovector pieces, and a determinantal term, the 't Hooft
interaction, which breaks the $U_A(1)$ symmetry. We use the following
Lagrangian: 
\begin{equation}
\begin{array}{rcl}
{\cal L\,} & = & \bar q\,(\,i\, {\gamma}^{\mu}\,\partial_\mu\,-\,\hat m)\, q+%
\frac{1}{2}\,g_S\,\,\sum_{a=0}^8\, [\,{(\,\bar q\,\lambda^a\, q\,)}%
^2\,\,+\,\,{(\,\bar q \,i\,\gamma_5\,\lambda^a\, q\,)}^2\,] \\[4pt] 
& - & \frac{1}{2}\,g_V\,\,\sum_{a=0}^8\, [\,{(\,\bar q\,\gamma_\mu\,%
\lambda^a\, q\,)}^2\,\,+\,\,{(\,\bar q \gamma_\mu\,\gamma_5\,\lambda^a\, q\,)%
}^2\,] \\[4pt] 
& + & g_D\,\, \{\mbox{det}\,[\bar q\,(\,1\,+\,\gamma_5\,)\,q\,] + \mbox{det}%
\,[\bar q\,(\,1\,-\,\gamma_5\,)\,q \,]\, \} \label{1} \\ 
&  & 
\end{array}
\label{eq:lag}
\end{equation}
In order to discuss the predictions of different models we consider the
cases: $g_V=0$ with two parametrizations (NJL I and NJL II) and $g_V\neq 0$
(ENJL). The model parameters, the bare quark masses $m_d=m_u, m_s$, the
coupling constants and the cutoff in three-momentum space, $\Lambda$, are
essentially fitted to the experimental values of $m_\pi,\,f_\pi,\,m_K$ and
to the phenomenological values of the quark condensates, $<\bar uu>,\, <\bar 
dd>,\,<\bar ss>$. The parameter sets used are, for NJL I: $\Lambda=631.4$
MeV, $g_S\,\Lambda^2=3.658,\,g_D\,\Lambda^5=- 9.40, \, m_u=m_d=5.5$ MeV and $%
m_s=132.9$ MeV; for ENJL: $\Lambda=750$ MeV, $g_S\,\Lambda^2=3.624$, $%
g_D\,\Lambda^5=- 9.11,\, g_V\,\Lambda^2=3.842$, $m_u=m_d=3.61$ MeV and $%
m_s=88$ MeV. For NJL II we use the parametrization of \cite{RKH}, $%
\Lambda=602.3$ MeV, $g_S\,\Lambda^2=3.67$, $g_D\Lambda^5=-12.39$, $%
m_u=m_d=5.5$ MeV and $m_s=140.7$ MeV, which underestimates the pion mass ($%
m_\pi\,=135\mbox{ MeV}$) and of $\eta$ by about $6\%$. \vskip0.5cm

The six quark interaction can be put in a form suitable to use the
bosonization procedure (see \cite{Vogl,Ripka,Ruivo}):

\begin{equation}
{\cal L_D}\,=\, \frac{1}{6} g_D\,\, D_{abc} \,(\bar q\, {\lambda}%
^c\,q\,)\,[\,(\,\bar q\,\lambda^a\, q\,)(\bar q\,\lambda^b\, q\,) - 3\,(\,%
\bar q \,i\,\gamma_5\,\lambda^a\, q\,)\,(\,\bar q \,i\,\gamma_5\,\lambda^b\,
q\,)\,]
\end{equation}

\noindent with: $D_{abc}=d_{abc}\,, a,b,c\,\, \epsilon\, \{1,2,..8\}\,, %
\mbox{(structure constants of SU(3))}\,,D_{000}=\sqrt{\frac{2}{3}}\,,
D_{0ab}=-\sqrt{\frac{1}{6}}\delta_{ab}$.

The usual procedure to obtain a four quark effective interaction from this
six quark interaction is to contract one bilinear $(\bar q\,\lambda_a\,q)$.
Then, from the two previous equations, an effective Lagrangian is obtained:

\vspace{1cm}

\begin{eqnarray}
L_{eff}\,&=& \bar q\,(\,i\, {\gamma}^{\mu}\,\partial_\mu\,-\,\hat m)\, q \,\,
\nonumber \\
&+& S_{ab}[\,(\,\bar q\,\lambda^a\, q\,)(\bar q\,\lambda^b\, q\,)]
+\,P_{ab}[(\,\bar q \,i\,\gamma_5\,\lambda^a\, q\,)\,(\,\bar q
\,i\,\gamma_5\,\lambda^b\, q\,)\,]  \nonumber \\
&-&\frac{1}{2}\,g_V\,\,\sum_{a=0}^8\, [\,{(\,\bar q\,\gamma_\mu\,\lambda^a\,
q\,)}^2\,\,+\,\,{(\,\bar q \gamma_\mu\,\gamma_5\,\lambda^a\, q\,)}^2\,]\,\,
\end{eqnarray}

\noindent where: 
\begin{eqnarray}
S_{ab}\,=\,g_S\,\delta_{ab}\,+\,g_D D_{abc} \,< \bar q\, {\lambda}^c\,q\,> 
\nonumber \\
P_{ab}\,=\,g_S\,\delta_{ab}\,-\,g_D D_{abc} \,< \bar q\, {\lambda}^c\,q\,>
\end{eqnarray}

By using the usual methods of bosonization one gets the following effective
action: 
\begin{eqnarray}  \label{action}
I_{eff}=&-i&Tr\ {\rm ln}(\,i\,\partial_\mu \gamma_{\mu}-\hat m%
+\sigma_a\,\lambda^a +i\,\gamma_5\, {\phi}_a\,\,{\lambda}^a\,+\,\gamma^\mu
\,V_\mu\,+\gamma_5\,\gamma^\mu\,A_\mu)  \nonumber \\
&-&\frac{1}{2}(\,\sigma_a\,{S_{ab}}^{-1}\,\sigma_b\,+{\phi}_a\, {\ P_{ab}}%
^{-1}\,\phi_b\,)  \nonumber \\
&+&\frac{1}{2G_V}(\,{V^a_{\mu}}^2\,+\,{A^a_{\mu}}^2\,),
\end{eqnarray}

\noindent from which we obtain the gap equations and meson propagators.

In order to introduce the finite temperature and density, we use the thermal
Green function, which, for a quark $q_i$ at finite temperature $T$ and
chemical potential $\mu_i$ reads:

\begin{eqnarray}
S (\vec x -\vec x^{\prime},\tau -\tau^{\prime})& = &\frac{i}{\beta}{\sum}_n
e^{-i \omega_n (\tau - \tau^{\prime})}  \nonumber \\
& \int& \frac{d^3p}{(2 \pi)^3} \frac{e^{-i\,\vec p\,(\vec x -\vec x%
^{\prime})}}{\gamma_0 (i \omega_n + {\overline \mu_i}) - \vec \gamma . \vec p
- M_i}\,,
\end{eqnarray}

\noindent where $\beta =1/T\,, {\overline \mu_i}\,=\,\mu_i\,-\,\Delta E_i$, $%
\Delta E_i$ is the energy gap induced by the vector interaction, $M_i$ the
mass of the constituent quarks, $E_i\,=\,(p^2\,+\,M^2_i)^{1/2}$ and $%
\omega_n\,=\,(2 n + 1)\,\frac{\pi}{\beta}\,\,\,, n\,=\,0\,,\pm 1\,,\pm
2\,,....,$ are the Matsubara frequencies. The following gap equations are
obtained:

\begin{equation}
M_i\,=\,m_i\,-2\,g_S\,<\bar q_i\, q_i>\,-\,2\,g_D\,<\bar q_j\, q_j><\bar q%
_k\, q_k>
\end{equation}
\begin{equation}
\Delta E_i\,=\,2\,g_V\,<{q_i}^+\, q_i>
\end{equation}

\noindent with $i\,,j\,,k$ cyclic and $<\bar q_i\, q_i>$, $<{q_i}^+\, q_i>$
are respectively the quark condensates and the quark densities at finite $T$
and $\mu_i$.

The condition for the existence the poles in the propagators of kaons leads
to the following dispersion relation:

\begin{equation}
(\,1\,-\,K_P\,\,J_{PP}\,)\,\,(\,1\,-\,K_A\,\,J_{AA}\,)\,-\,K_P\,K_A\,\, {J^2}%
_{PA}\,=\,0
\end{equation}
with: 
\[
\omega\, J_{PA}\,=\,(M_u + M_s) \,J_{PP}\,+\,2\,(<\bar u u\,>+\,<\bar s s>). 
\]
\[
\omega\, J_{AA}\,=\,(M_u + M_s) \,J_{PA}\,+\,2\,(< u^+ u>\,-\,< s^+ s>).
\]

\[
J_{PP}\,=\,2\,N_c\,\int {\frac{d^3 p}{(2\pi)^3}}\,\left\{\frac {M_u (M_s -
M_u) \,- q_0\, E_u}{({E_s}^2 - (q_0+ E_u)^2) E_u} \mbox{ tanh} \frac{%
\beta(E_u+{\bar \mu_u)}}{2}\,+\right. 
\]
\begin{equation}
\left.\frac {M_u (M_s - M_u) \ + q_0 E_u}{({E_s}^2 - (q_0 -E_u)^2) E_u} %
\mbox{ tanh} \frac{\beta(E_u- {\bar \mu_u)}}{2}\,\,+\, s \rightarrow
u\,,\,q_0 \rightarrow -q_0\right\}
\end{equation}
with $K_P\,=\,g_S\,+\,g_D \,<\bar d d>$ and $K_A\,=\,-g_V$, ${\bar \mu_u}%
=\mu_u-\Delta E_u$, $q_0 =\pm m_{K^{\pm}}- (\Delta E_u-\Delta E_s)$ for $%
K^{\pm}$

Similar expressions are obtained for pions in neutron matter, by replacing $%
s\leftrightarrow d$ \cite{SousaRuivo}.
\section{Phase transitions at finite chemical potential and temperature}

We reanalyze the problem of the phase transitions in order to establish a
connection between the vacuum state and its excitations. We consider here
the case of asymmetric quark matter without strange quarks simulating
neutron matter: $\rho_u\,=\frac{1}{2}\,\rho_d\,\,,\rho_s\,=\,0$ and
calculate the the energy and pressure. At zero temperatures we found a first
order phase transition in NJL model, exhibiting different characteristics
according to the parameterization used. Within the parameterization NJL I,
the pressure is negative for $0.8 \rho_0\leq \rho \leq 1.65 \rho_0$ and the
absolute minimum of the energy per particle is at $\rho=0$ (dashed curves in
Fig.1).

\begin{figure}[h]
\epsfig{file=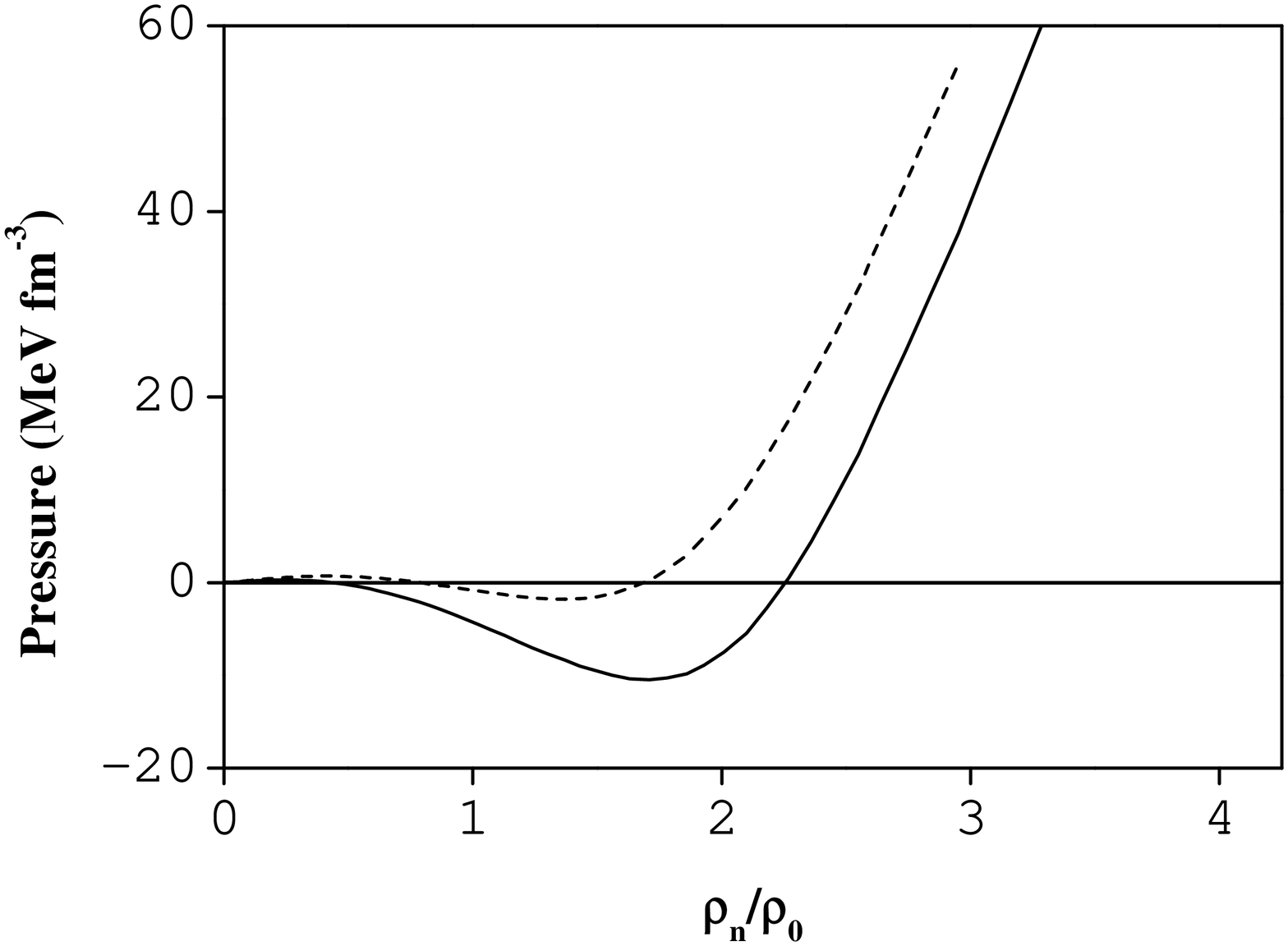,width=8.0cm,height=6.5cm}
\hfill
\epsfig{file=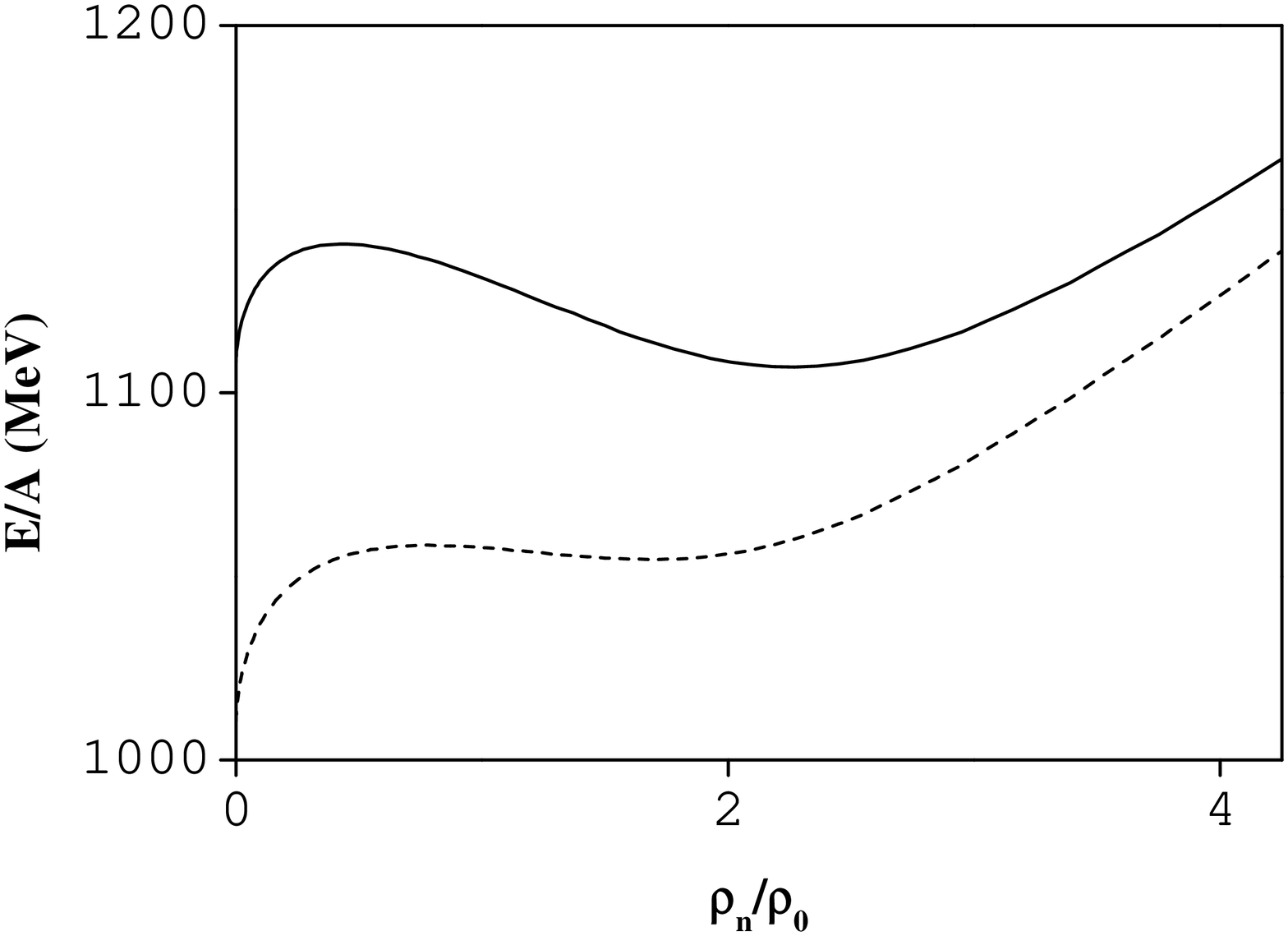,width=8.0cm,height=6.5cm}
\caption{Pressure and energy per particle in NJL I (dashed curves) and NJL
II (full curves).}
\end{figure}

The system, within the range of densities indicated above, is in a mixed
phase consisting of droplets of massive quarks of low density and droplets
of light quarks of high density and, for $\rho > 1.65 \rho_0$, is in a quark
phase with partially restored chiral symmetry (in the SU(2) sector). These
droplets are unstable since the absolute minimum of the energy per particle
is at $\rho=0$. With parameterization II, the mixed phase starts at $%
\rho\simeq 0$, because, although the zeros of the pressure are at $\rho =
0.44 \rho_0,\,\,\rho_c=2.25 \rho_0$, the compressibility is negative in the
low density region for $\rho \simeq 0$; the energy per particle has an
absolute minimum at the critical density of $E/A \,= 1102$ MeV, about three
times the masses of the constituent non strange quarks in vacuum (Fig 1.
full curves). The model may now be interpreted as having a hadronic phase
--- droplets of light $u\,,d$ quarks with a density $\rho_c=2.25 \rho_0$
surrounded by a non trivial vacuum --- and, above the critical density, a
quark phase with partially restored SU(2) chiral symmetry. The model is not
suitable to describe hadrons for $\rho\,< \rho_c$. Since there is no
definition of the density in the mixed phase, we study, in the following,
the mesonic excitations only for $\rho > \rho_c$.

The phase transition becomes second order at finite density and temperatures
around $20 MeV$ and also in the  ENJL model, for the set of parameters
chosen. In these cases the system has positive pressure but the absolute
minimum of the energy per particle is at zero density. Although a gas of
quarks does not exist at low densities, the model has been used to study the
influence of the medium in the mesonic excitations of the vacuum. Of course,
an extrapolation of quark matter to hadronic matter should be made, since we
do not have, at low densities, a gas of hadrons.

\section{Behavior of pions and kaons in the medium}

At zero temperature we observe a splitting between charge multiplets, both
in NJL and ENJL models (see figs. 2-3).

\begin{figure}[h]
\epsfig{file=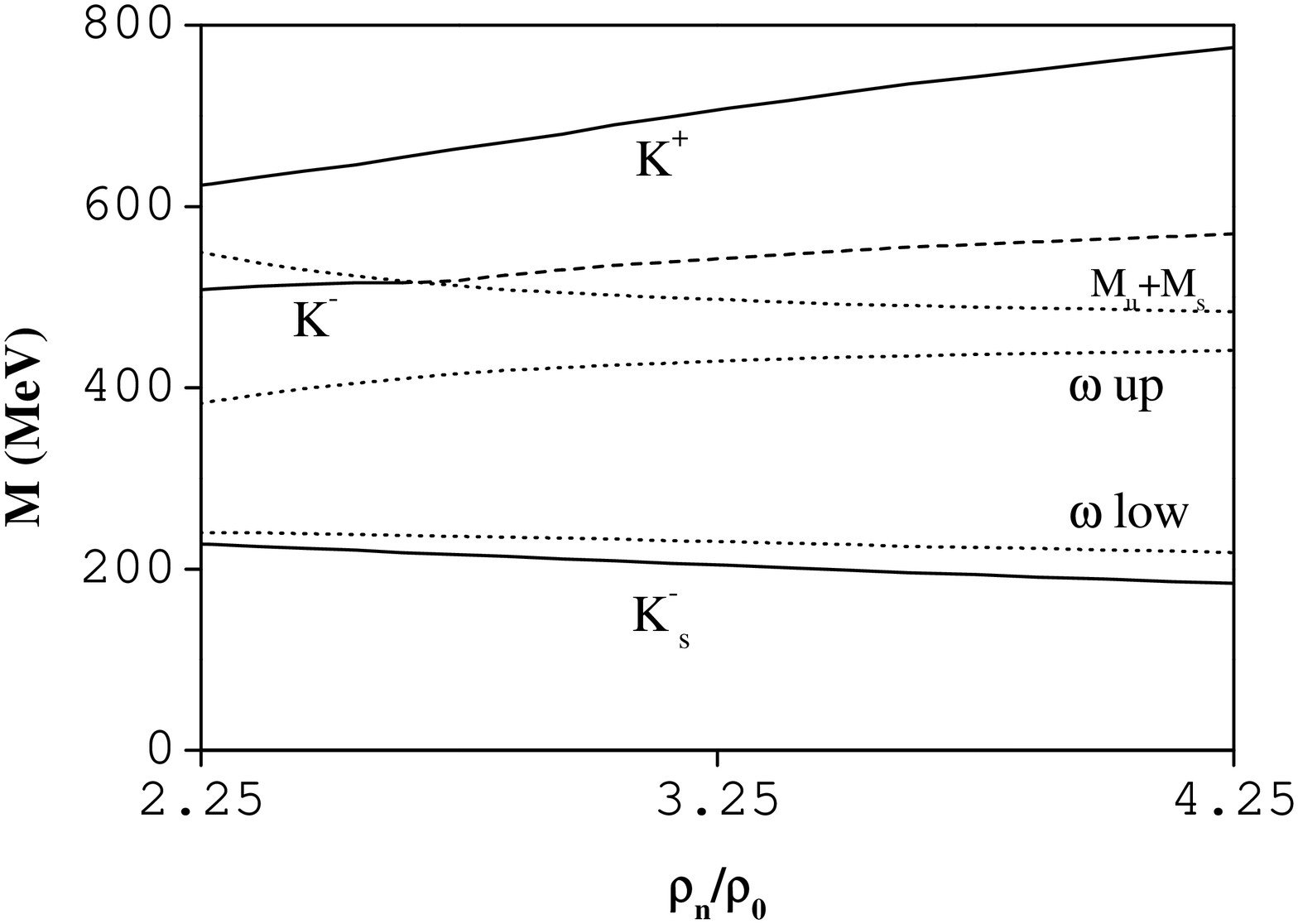,width=8.0cm,height=6.5cm}
\hfill
\epsfig{file=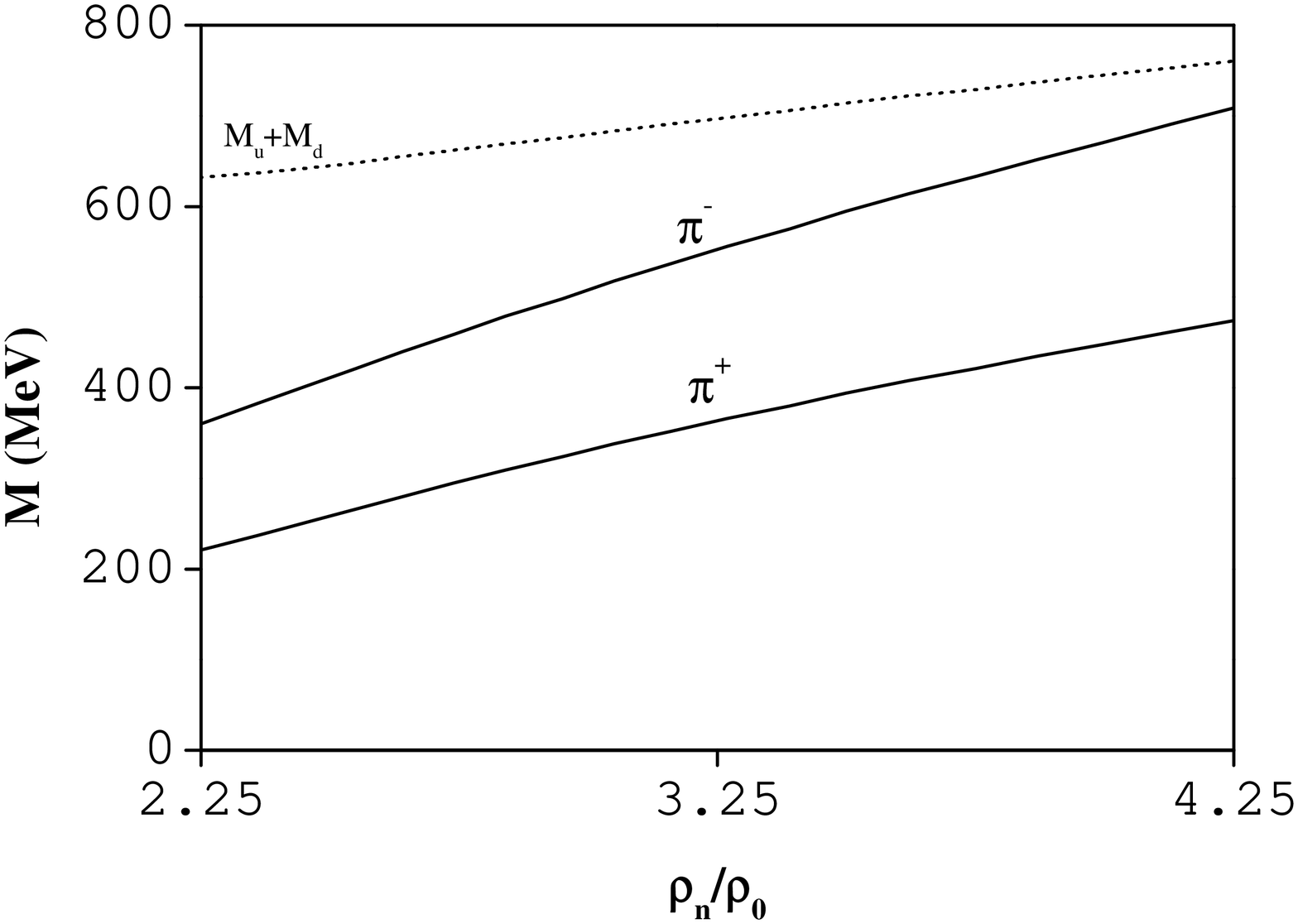,width=8.0cm,height=6.5cm}
\caption{Masses of kaons and pions in NJL II. $\protect\omega_{up}\, 
\mbox{and }\, \protect\omega_{low}$ denote the limits of the Fermi sea
continuum.}
\end{figure}

These modes are excitations of the Dirac sea modified by the presence of the
medium. The increase of $K^+ \mbox{and}\, \pi^- $ masses with respect to
those of $K^- \, \mbox{and}\, \pi^+$ is due to Fermi blocking and is more
pronounced for kaons than for pions because, there are $u$ and $d$ quarks in
the Fermi sea and therefore there are repulsive effects due to the Pauli
principle acting on $\pi^+$. In the case of kaons, we do not have strange
quarks at these densities and so there is mora phase space available to
create $\bar u s$ pairs of quarks and there are less repulsive effects on $%
K^-$.

An interesting feature in NJL model is that, besides these modes, low energy
modes with quantum numbers of $K^- \, \mbox{and}\, \pi^+$ appear. These are
particle-hole excitations of the Fermi sea which correspond to $\Lambda\,
(1106)$-particle-proton-hole for kaons an to a proton-particle-neutron-hole
for the case of pions. A similar effect is found for kaons in symmetric
nuclear matter \cite{SousaRuivo,RSP}. For the case of pions, the low energy
modes is less relevant and exist only for $\rho=\rho_c$, merging in the
Fermi sea continuum afterwards. We notice that when the 't Hooft interaction
is not included \cite{SousaRuivo} or in SU(2) \cite{Hiller} the low energy
mode for pions is more relevant.

The sum rules are a very important tool to analyze the collectivity and
relative importance of the modes \cite{RuivoSousa,SousaRuivo,Hiller,RSP}.
One can derive a generalization of the PCAC relation in the medium from the
Energy Weighted Sum Rule (EWSR), well known from Many Body Theories. For the
mesonic state $|r>$ with energy $\omega_r$ associated with the transition
operator $\Gamma$  the strength function $F_r\,=\,\omega_r\,\,{%
|\,<\,r\,|\,\Gamma\, |\,0\,>\,|}^2$ satisfies the EWSR \, which reads 
\begin{equation}
m_1\,=\,{\sum}_r\,\omega_r\,\,{|\,<\,r\,|\,\Gamma\, |\,0\,>\,|}^2\,\,=\, 
\frac{1}{2}\, <\,\Phi_0\,|\,[\,\Gamma\,,\,[\,H\,,\Gamma\,]\,]\,|\, \Phi_0\,>,
\label{10}
\end{equation}
the transition operator being defined in the present case by $%
\Gamma\,=\,\Gamma_+\,+\,\Gamma_-$, with $\Gamma_\pm\,=\gamma_5\,(\lambda_4%
\pm i\ \lambda_5)/\sqrt{2}$, for kaons, and , $\Gamma_\pm\,=\gamma_5\,(%
\lambda_1\pm i\ \lambda_2)/\sqrt{2}$, for pions. We obtain therefore the
GMOR relation in the medium:

\begin{equation}
\sum_\alpha \,m_{K,\alpha}^2\,f_{K,\alpha}^2\,\,\simeq\,-\,\frac{1}{2}
(m_u\,+\,m_s\,)\,[<\,\overline u\,\, u\,>\,+\, <\,\overline s\,\,s\,>\,]\,.
\label{11}
\end{equation}
and 
\begin{equation}
\sum_\alpha \,m_{\pi,\alpha}^2\,f_{\pi,\alpha}^2\,\,\simeq\,-\,\frac{1}{2}
(m_u\,+\,m_d\,)\,[<\,\overline u\,\, u\,>\,+\, <\,\overline d\,\,d\,>\,]\,.
\label{12}
\end{equation}

We verified that in the medium the degree of satisfaction of the sum rule is
good, provided, naturally, that all the bound state solutions are
considered. The strength associated to the low energy mode can not be
neglected as the density increases \cite{RuivoSousa,SousaRuivo}.

\begin{figure}[h]
\epsfig{file=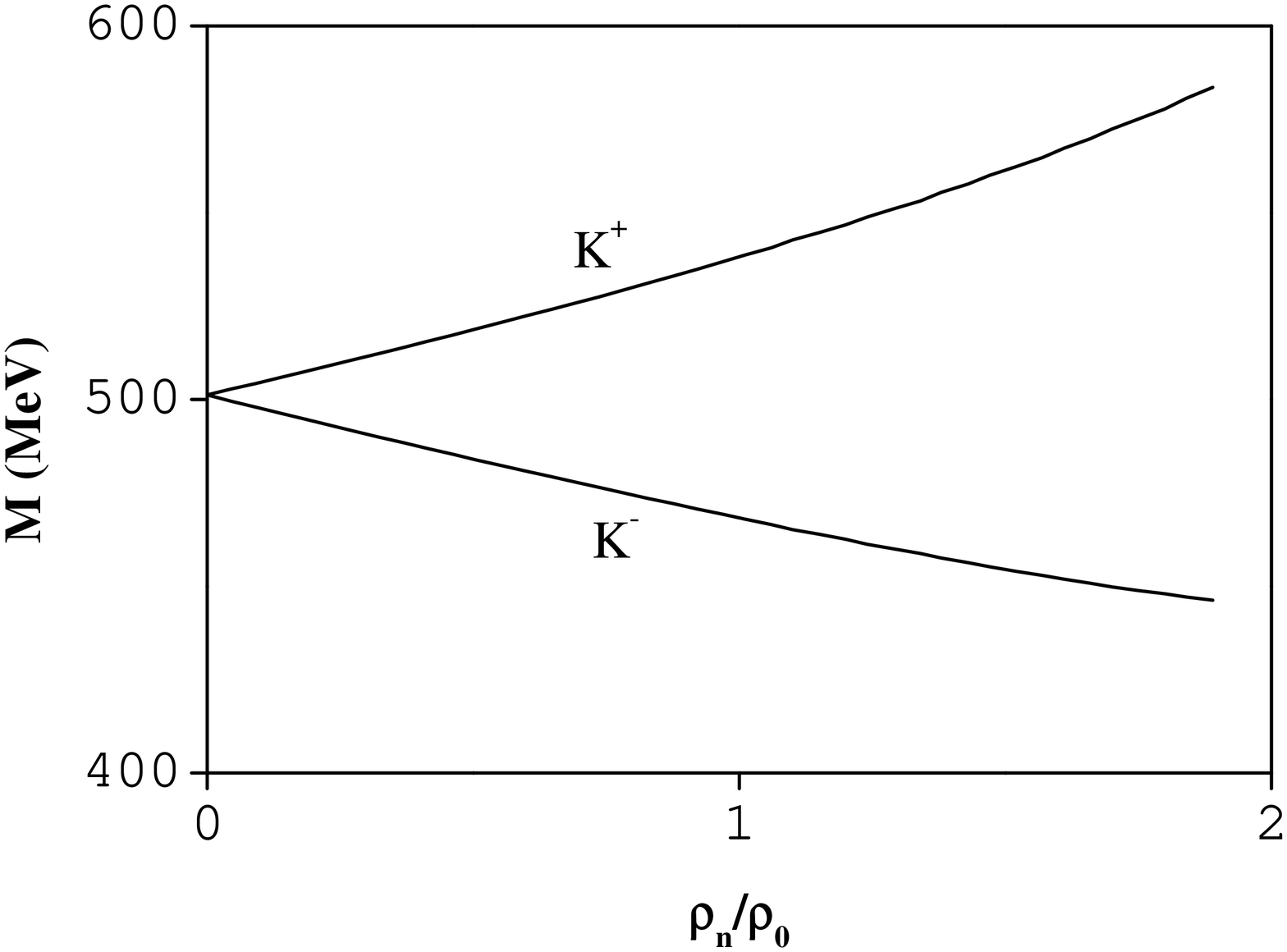,width=8.0cm,height=6.5cm}
\hfill
\epsfig{file=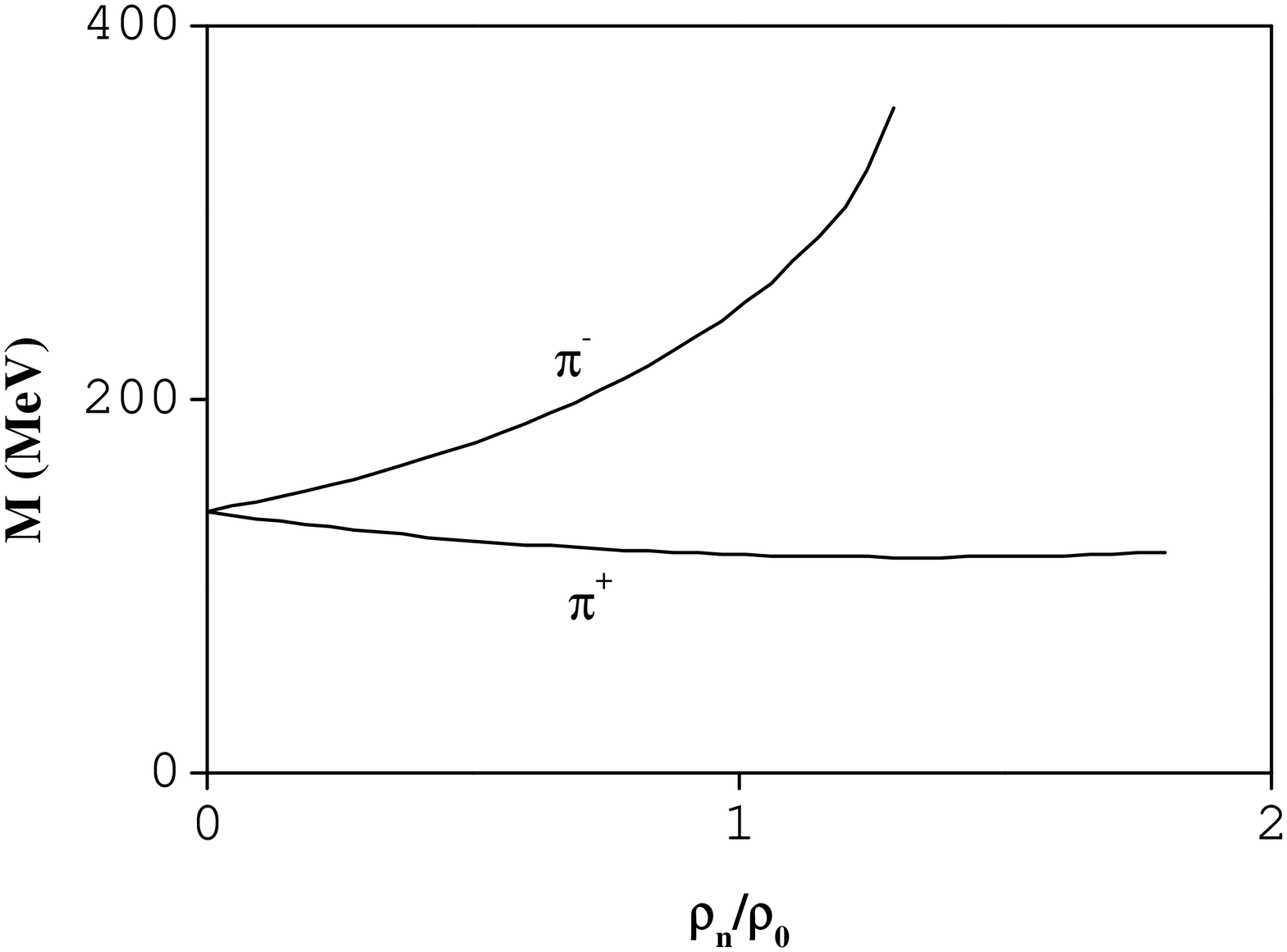,width=8.0cm,height=6.5cm}
\caption{Masses of kaons and pions in ENJL model.}
\end{figure}

In the ENJL model the low energy mode does not appear, which is consistent
with the fact that the analysis of the EOS shows that there is no stable
Fermi sea. The attractive effects concentrate on $K^-\,\,,\pi^+$. The
splitting between the charge multiplets is even larger in this model.

The influence of the temperature is, on one side, to inhibit the occurrence
of the low energy mode (this mode is not seen above very low temperatures)
and, on the other side, to reduce the splitting of between the upper energy
modes (see Fig. 4).

We notice that temperature has an effect on the low energy mode similar to
the vector pseudovector interaction in vacuum. This is meaningful since, as
it has been mentioned above, as the temperature increases the phase
transition is second order and the minimum of energy per particle is at $%
\rho=0$ and consequently the Fermi sea is not stable. So, it is reasonable
that the excitations of the Fermi sea are not seen.

\begin{figure}[h]
\epsfig{file=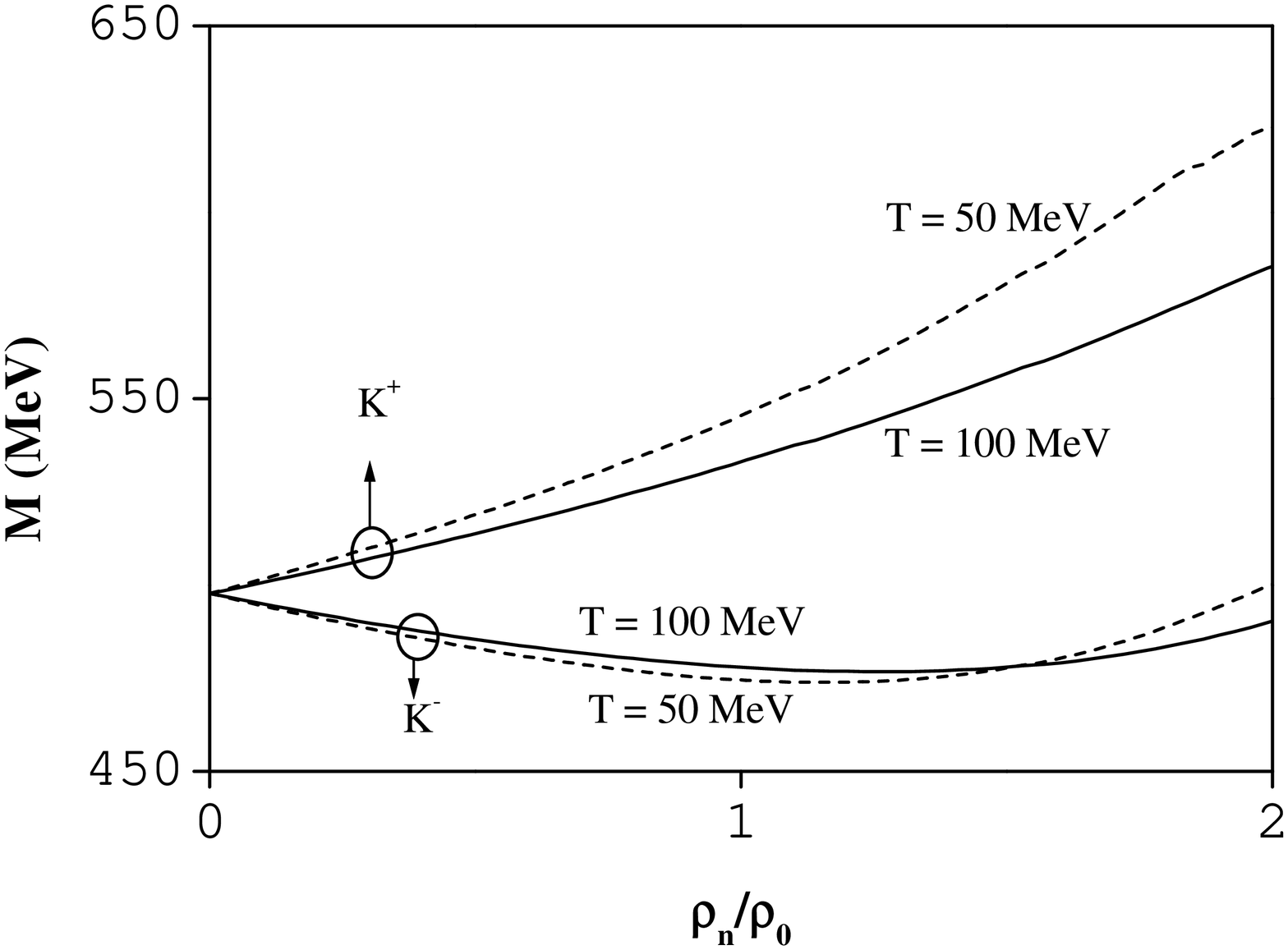,width=8.0cm,height=6.5cm}
\hfill
\epsfig{file=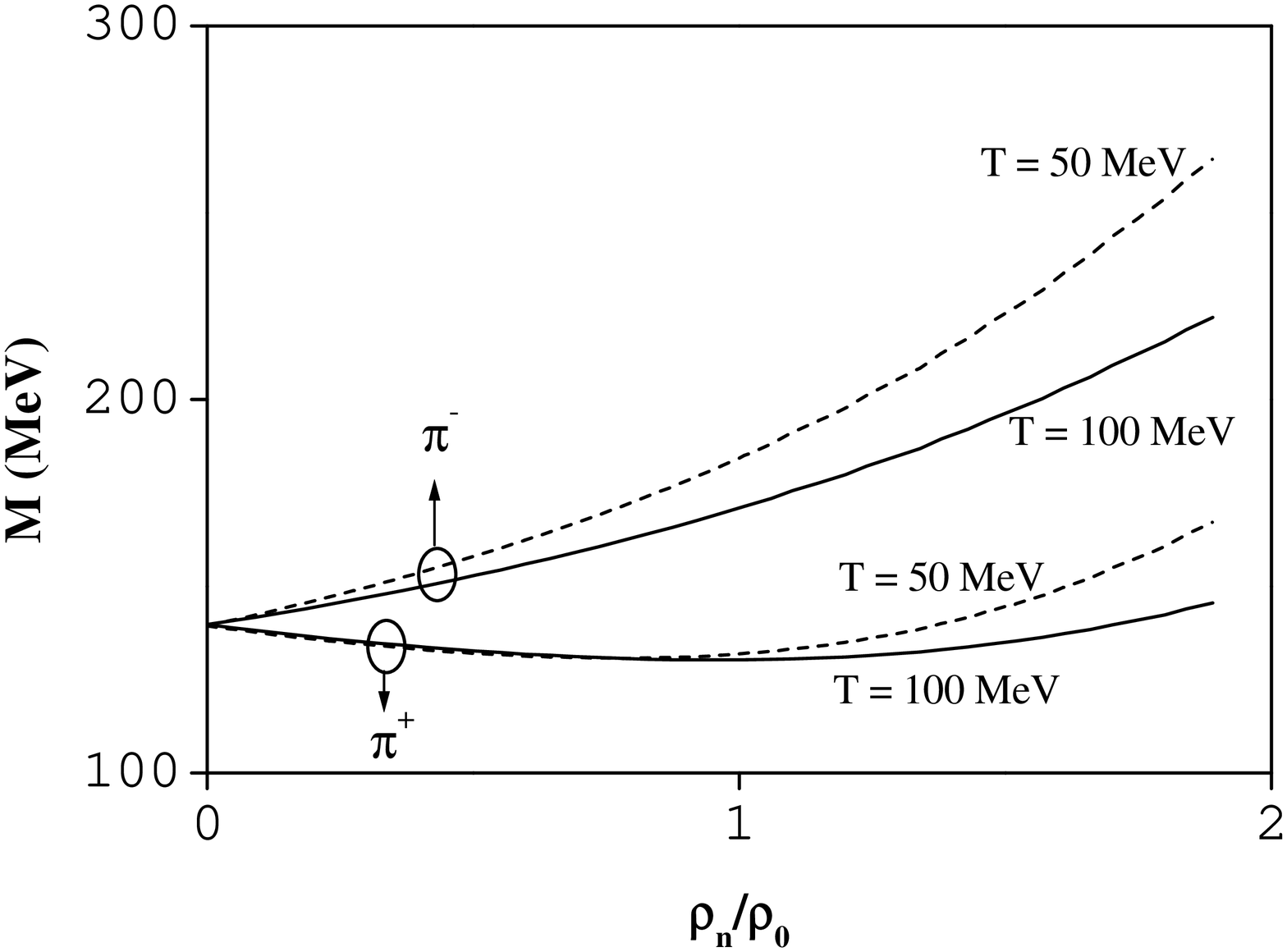,width=8.0cm,height=6.5cm}
\caption{Masses of kaons and pions for $T=50$ MeV and $T=100$ MeV, in NJL II. 
}
\end{figure}

In conclusion, we have discussed the behavior of kaons and pions in neutron
matter in NJL and ENJL model, in connection with the nature of the phase
transition, at finite density with zero or non zero temperature. In the NJL
model where the phase transition is first order and a stable Fermi sea
exists at the critical density, we find low energy particle hole excitations
of the Fermi sea, besides the usual splitting of charge multiplets which are
excitations of the Dirac sea. This last effect does not occur in the ENJL
model, where the transition is second order. The temperature inhibits this
effect and reduces the splitting between the charge multiplets.

\end{document}